\RequirePackage[T1]{fontenc}
\documentclass[journal]{IEEEtran}

\usepackage{cite}
\usepackage{amsmath}
\usepackage{graphicx}
\usepackage{booktabs}
\usepackage{array}
\usepackage{xurl}
\usepackage{iftex}
\usepackage[hidelinks]{hyperref}

\ifPDFTeX
\else
  \ifXeTeX
    
  \fi
\fi

\newcolumntype{L}[1]{>{\raggedright\arraybackslash}p{#1}}
\title{CrowdStrike Channel File 291: An Engineering Analysis of a Global Endpoint Outage}
\author{Luiz~Parente\\\textit{lparente@purdue.edu}%
\thanks{Luiz Parente is with Purdue University, West Lafayette, IN, USA
(e-mail: lparente@purdue.edu).}}

\hypersetup{
  pdftitle={CrowdStrike Channel File 291: An Engineering Analysis of a Global Endpoint Outage},
  pdfauthor={Luiz Parente},
  pdfsubject={Engineering incident analysis},
  pdfkeywords={configuration management, dependability, endpoint security, incident analysis, software deployment}
}

\begin{document}
\maketitle

\begin{abstract}
On July 19, 2024, a CrowdStrike Falcon content update caused Windows systems
to crash, affecting about 8.5 million devices according to a Microsoft
estimate. The immediate defect
was an inconsistency between a detection template that declared 21 inputs
and sensor code that supplied 20. A new matching criterion exposed an
unchecked array access in the kernel-mode content interpreter. This article
reconstructs the incident from vendor technical reports, Microsoft's crash
analysis and recovery guidance, congressional records, and operational
impact reports. It distinguishes documented findings from engineering
inferences and examines how interface validation, test selection, deployment
policy, and recovery dependencies interacted. Four assurance gaps were
consequential: validation against a definition the consumer did not meet,
tests that avoided the missing input, distribution without health-gated
promotion, and recovery paths requiring connectivity that some crashing
hosts could not maintain. Recommended controls separate
defect prevention, exposure containment, and service restoration, with
verification criteria, release-cadence constraints, and operational
tradeoffs. The case shows why configuration consumed by privileged software
requires lifecycle assurance comparable to code, and why release reversion,
endpoint recovery, and service restoration must be assessed separately.
\end{abstract}


\section{Introduction}
\label{sec:introduction}

Endpoint security software occupies an unusually consequential position
in an enterprise system. It must observe activity that ordinary
applications cannot reliably inspect, resist tampering, and receive
frequent updates as threats change. Those capabilities also make the
security product part of the availability boundary of every host on which
it runs. A defective update can disable the systems it is intended to
protect.

The CrowdStrike outage of July 19, 2024, illustrates this dependency. A
Rapid Response Content update delivered through Channel File 291 activated
a latent defect in the Falcon sensor for Windows. The affected detection
template declared 21 input fields, whereas the sensor integration supplied
20. Earlier content used a wildcard for the final field; the new content
required that field to be inspected. The resulting out-of-bounds read
occurred in kernel mode and caused a Windows bugcheck
\cite{crowdstrike_rca}. Microsoft estimated that 8.5 million Windows devices
were affected, fewer than 1\% of the Windows population, while noting that
their concentration in enterprises operating critical services amplified
the disruption \cite{microsoft_weston_blog}.

The immediate memory-safety defect does not fully explain the incident's
scale or recovery burden. The incompatible interface survived development
and testing, content validation accepted the triggering instance, and the
update was distributed without a staged rollout capable of limiting initial
exposure. Reverting the update then addressed distribution, but did not
necessarily restore hosts that could no longer boot normally. Each stage
requires a different engineering control.

This article examines the incident as a retrospective engineering case
study. Its purpose is to connect the published technical explanation to
release assurance and operational recovery, rather than to introduce a new
detection method or claim an experimental result. The central question is:
how did a bounded input-contract defect become a cross-organizational
availability incident, and which controls would address the distinct stages
of that escalation?

The analysis contributes a source-traceable reconstruction, a distinction
between shared and independent defensive assumptions, and a set of
verification criteria for preventing, containing, and recovering from
similar failures. Section~\ref{sec:method} establishes the evidence and
analytical method. Section~\ref{sec:reconstruction} reconstructs the system,
failure, and operational consequences. Sections~\ref{sec:causal} and
\ref{sec:controls} examine the causal structure and corresponding controls.
Section~\ref{sec:discussion} addresses responsibility, standards, and the
limits of the conclusions.

\section{Evidence and Analytical Method}
\label{sec:method}

\subsection{Scope and Source Selection}

The system boundary includes the cloud-side content pipeline, the Windows
sensor's interpretation of that content, its deployment to customer
fleets, and the recovery paths available after a host failure. The incident
record considered here extends from the introduction of the affected
template in February 2024 through public response and resilience
assessments published by October 31, 2024. Sources were checked on
October 22, 2026. This is a reconstruction of the 2024 incident and response,
not an assessment of the current Falcon product.

Sources were selected for direct relevance to mechanism, deployment,
recovery, or operational impact. CrowdStrike's August 6 root-cause analysis
(RCA) provides the most detailed account of the input mismatch and test
coverage \cite{crowdstrike_rca}. Its preliminary post-incident review
(PIR) and technical update report the release process and event times
\cite{crowdstrike_pir,crowdstrike_techdetails}. Microsoft's published crash
analysis supplies a separate operating-system perspective on the fault
\cite{microsoft_security_best_practices}; its recovery guidance documents
practical restoration options \cite{microsoft_recovery_tool}.
CrowdStrike's remediation hub supplies dated recovery guidance and
vendor-reported sensor-connection measurements
\cite{crowdstrike_recovery_hub}.

For deployment policy and impact, the analysis uses the House Homeland
Security Committee's account of September 2024 testimony and Meyers'
written statement, a Congressional Research Service (CRS) report, Delta
Air Lines' quarterly results, and the UK Financial Conduct Authority's
(FCA's) retrospective
\cite{homeland_hearing,meyers_testimony,congress_crs,delta_results,fca_resilience}.
CrowdStrike's October 25 complaint against Delta documents a competing
causal attribution, not an adjudicated finding
\cite{crowdstrike_delta_complaint}.
The Cybersecurity and Infrastructure Security Agency (CISA) advisory
provides incident-response context \cite{cisa_alert}. These sources were
chosen purposively; the article does not claim an exhaustive literature
review, a census of affected organizations, or a complete loss estimate.

\subsection{Evidence Handling}

The reconstruction separates three kinds of statement. A
\emph{reported finding} describes what a cited source establishes within its
visibility. A \emph{corroborated observation} is supported by a separate
technical or operational record. An \emph{engineering inference} connects
those observations or proposes a control; it is not presented as a fact
about CrowdStrike's internal decisions. Table~\ref{tab:evidence} applies
this classification to the principal incident claims and their limits.

\begin{table*}[t]
\caption{Principal Incident Claims and Their Evidentiary Limits}
\label{tab:evidence}
\centering
\small
\begin{tabular}{@{}L{0.27\textwidth}L{0.30\textwidth}L{0.37\textwidth}@{}}
\toprule
Claim & Primary support & Interpretation and limit \\
\midrule
The template declared 21 fields; the sensor supplied 20; a non-wildcard
criterion exposed the mismatch.
& RCA, pp.~2--5 and 10--12 \cite{crowdstrike_rca}.
& Reported finding (vendor). No sensor source code or original
crash dump was examined for this article. \\[3pt]
A kernel-mode invalid read in \texttt{csagent.sys} caused the observed
bugcheck.
& Microsoft's crash analysis \cite{microsoft_security_best_practices};
RCA, pp.~8--12 \cite{crowdstrike_rca}.
& Corroborated observation. Microsoft confirms the fault, not the complete
21-versus-20 interface history. \\[3pt]
Testing did not exercise a non-wildcard criterion in field 21.
& RCA findings 1, 3, and 5 \cite{crowdstrike_rca}.
& Reported finding (vendor). Template Type tests ran on supported Windows
versions without forcing a field-21 read; per-instance execution coverage
was incomplete. \\[3pt]
Content distribution lacked a staged rollout.
& RCA finding 6 \cite{crowdstrike_rca}; committee account of Meyers'
testimony \cite{homeland_hearing}.
& Reported finding (vendor). The RCA and testimony repeat the vendor's
account; publication in two venues does not make it independent. \\[3pt]
Distribution reversion and endpoint restoration were different operations.
& Release timestamps, vendor remediation guidance, and Microsoft's repair
procedures
\cite{crowdstrike_techdetails,crowdstrike_recovery_hub,microsoft_recovery_tool}.
& Reported finding (vendor). Restoration paths depended on connectivity,
encryption, and tooling; a physical visit was not universal. \\[3pt]
Approximately 8.5 million Windows devices, fewer than 1\% of all Windows
machines, were affected.
& Microsoft's July 20 estimate \cite{microsoft_weston_blog}.
& Reported finding (Microsoft estimate). No method or device-level recovery
dataset is supplied; CISA's July 21 relay repeats rather than corroborates
the estimate \cite{cisa_alert}. \\[3pt]
Windows sensor connections were approximately 99\% of the pre-incident
comparison level by July 29.
& Vendor measurement published July 31, 16:38 UTC
\cite{crowdstrike_recovery_hub}.
& Reported finding (vendor metric). Week-over-week connection proxy with
approximately 1\% normal variance; it cannot give a recovery percentage
for the 8.5 million affected devices or for services. \\
\bottomrule
\end{tabular}
\end{table*}

The RCA adds detail to the PIR's initial description of a validator bug,
while retaining the characterization ``logic error'' in finding 4
\cite[p.~5]{crowdstrike_rca}. This article interprets the reported behavior
as an agreement failure: content could conform to the declared 21-field
template while remaining incompatible with the sensor's 20-input
implementation. That interpretation does not establish that the validator
was otherwise correct; calling the content simply ``malformed'' would
obscure the interface mismatch. Meyers' later written statement describes
an extra content input without a defined action \cite[p.~4]{meyers_testimony};
the reconstruction uses the RCA's more precise distinction between
declared and supplied inputs.

The analytical procedure follows the content from definition to
validation, deployment, interpretation, and recovery. At each transition,
it identifies the required invariant, the documented failure, and the
control capable of interrupting that failure. The barrier and fishbone
diagrams organize these relationships; they do not estimate causal
probabilities or establish employees' intentions.

\subsection{Relationship to Established Practice}

Chen et al.\ documented configuration dependencies within and across
components in 16 cloud and datacenter systems \cite{chen2020_config_dependencies}.
Their work supports treating interface dependencies as an explicit
engineering concern. It does not establish the behavior of the Falcon
implementation or show that a particular schema language would have
prevented this incident. Google's guidance on canary releases provides
established practice for evaluating a change on a limited population before
wider deployment \cite{google_canary}. Its assumptions about observation
and rollback require particular care when the changed component can crash
the host that supplies those signals.

Reason's system-oriented account of failure provides the basis for
examining defenses rather than reducing the event to an individual's
mistake \cite{reason2000}. Here that perspective is used as a qualitative
organizing device. The article does not treat a visual alignment of
barriers as proof that all barriers are independent or that any one
intervention would guarantee prevention.

\section{System and Incident Reconstruction}
\label{sec:reconstruction}

\subsection{Two Update Lifecycles, One Runtime Consumer}

The Falcon sensor is an endpoint detection and response (EDR) component.
On Windows, its kernel-mode components provide visibility and enforcement
for security-relevant activity. Microsoft's analysis identifies
\texttt{csagent.sys} as a file system filter driver and describes the
named-pipe notifications relevant to the incident
\cite{microsoft_security_best_practices}. Execution in the kernel makes a
sensor fault potentially consequential for the operating system, rather
than only for an isolated application process.

CrowdStrike distinguished two update lifecycles. \emph{Sensor Content}
ships with a sensor release and includes compiled Template Types. The
reported release process included automated and manual testing, internal
deployment, early adopters, and customer-selected sensor version policies.
\emph{Rapid Response Content} consists of matching criteria delivered as
Template Instances in Channel Files. It can change detection behavior
without distributing a new sensor binary \cite{crowdstrike_pir}.
Channel Files use a \texttt{.sys} extension on Windows, but are configuration
data, not kernel drivers \cite{crowdstrike_techdetails}.

A Template Type defines a category of inputs; a Template Instance supplies
matching criteria for those inputs. The Template Type Definitions file
describes the fields against which the cloud-side Content Validator checks
content. On the endpoint, sensor integration code supplies activity data
to the Content Interpreter, which evaluates the criteria
\cite[pp.~1--2, 7]{crowdstrike_rca}. Correctness therefore depends on
agreement among the definition, the delivered instance, and the actual
inputs supplied by every compatible sensor version.

\begin{figure*}[p]
\centering
\includegraphics[width=0.96\textwidth,height=0.90\textheight,keepaspectratio]{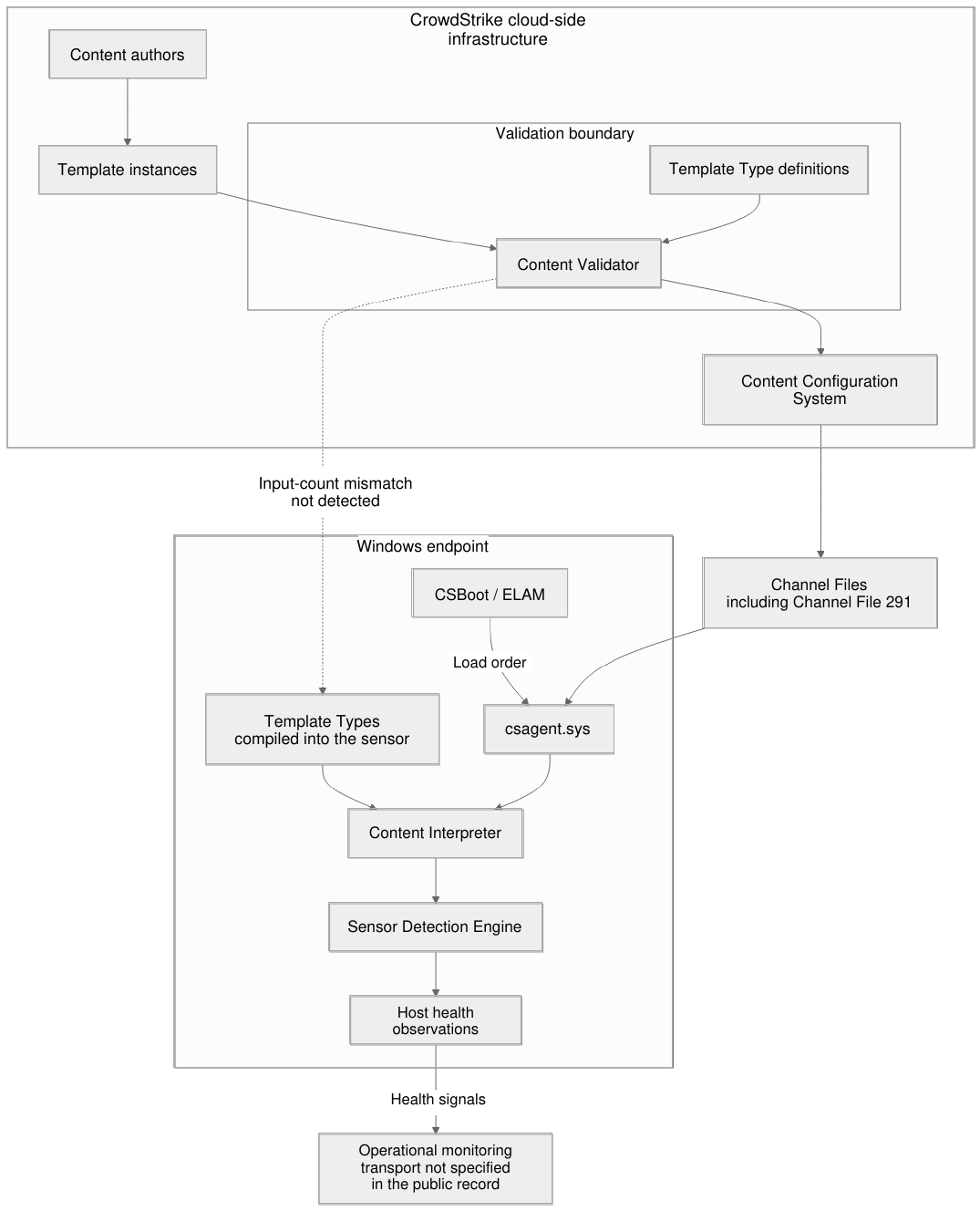}
\caption{Functional reconstruction of the content pipeline, based on
\cite{crowdstrike_pir,crowdstrike_rca,microsoft_security_best_practices}.
The validation boundary did not detect disagreement with the sensor's
actual input count. The monitoring path denotes a functional feedback
requirement, not a verified internal transport or a guarantee of crash
report delivery. ELAM denotes Early Launch Antimalware.}
\label{fig:arch}
\end{figure*}

\begin{figure*}[t]
\centering
\includegraphics[width=\textwidth]{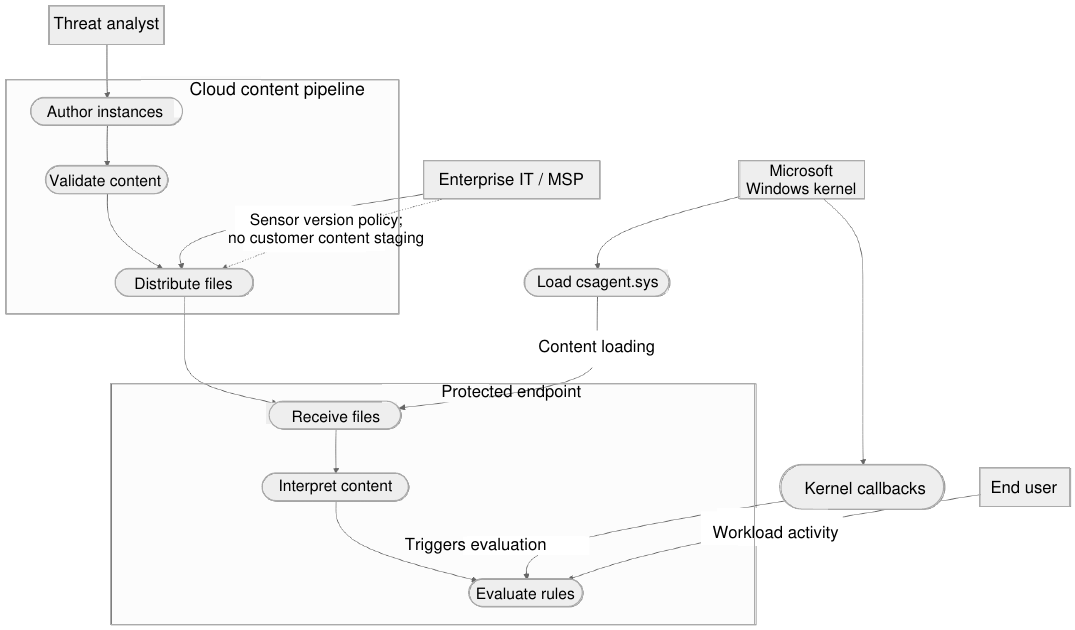}
\caption{Actor and responsibility view of the pre-incident content path
\cite{crowdstrike_pir,homeland_hearing}. Customers could select sensor
versions; this did not provide an equivalent staging policy for Rapid
Response Content. Arrows summarize responsibilities and dependencies,
rather than a complete execution sequence. MSP denotes managed service
provider.}
\label{fig:usecase}
\end{figure*}

Figure~\ref{fig:arch} summarizes these functional boundaries. It is not a
complete reconstruction of proprietary service topology. In particular,
the monitoring path is conceptual: a crash does not guarantee that the
failing sensor can deliver a health report. Figure~\ref{fig:usecase}
separates the vendor's content-release authority from customers' sensor
version selection and Microsoft's platform role.

\subsection{Timeline and Trigger}

Sensor version 7.11, released on February 28, 2024, introduced an
interprocess communication (IPC) Template Type for activity involving
Windows named pipes. The definition specified 21 input fields, while
the integration code supplied 20. The discrepancy was not detected
before release \cite{crowdstrike_rca}. Table~\ref{tab:timeline} separates
the introduction of this latent defect from its later activation.

\begin{table*}[t]
\caption{Incident and Reported Remediation Timeline in 2024.
Scheduled release dates are commitments reported in the August 6 RCA,
not independently verified completion dates.}
\label{tab:timeline}
\centering
\small
\begin{tabular}{@{}L{0.16\textwidth}L{0.78\textwidth}@{}}
\toprule
Date or time & Event \\
\midrule
Feb.~28 & Sensor 7.11 introduced the IPC Template Type
\cite{crowdstrike_pir}. \\[3pt]
Mar.~5 & Staging stress test passed; first IPC Template Instance
released \cite{crowdstrike_pir}. \\[3pt]
Apr.~8--24 & Three further IPC Template Instances were deployed;
field 21 remained a wildcard \cite{crowdstrike_pir,crowdstrike_rca}. \\[3pt]
Jul.~19,\newline 04:09 UTC & Two new instances were deployed;
one required non-wildcard matching in field 21
\cite{crowdstrike_pir,crowdstrike_rca}. \\[3pt]
Jul.~19,\newline 05:27 UTC & Problematic content was reverted,
78 minutes after release \cite{crowdstrike_techdetails}. \\[3pt]
Jul.~20--21 & Microsoft published its affected-device estimate
and recovery tooling \cite{microsoft_weston_blog,microsoft_recovery_tool}. \\[3pt]
Jul.~21 & CrowdStrike advised rebooting on a wired network to allow
replacement content to arrive before another crash; guidance updated
09:32 UTC \cite{crowdstrike_recovery_hub}. \\[3pt]
Jul.~22 & CrowdStrike introduced automated remediation techniques,
according to its later written testimony \cite{meyers_testimony}. \\[3pt]
Jul.~23 & Known-bad classification configured in US-1, US-2, and EU;
the July 24, 03:35 UTC hub update reports possible automatic recovery
with strong connectivity \cite{crowdstrike_recovery_hub}. \\[3pt]
Jul.~24 & PIR published, outlining testing, deployment, and
validation changes \cite{crowdstrike_pir}. \\[3pt]
Jul.~25 & Interpreter bounds and input-count checks added, according
to the RCA; sensor backports were still pending \cite{crowdstrike_rca}. \\[3pt]
Jul.~27 & Compile-time count validation entered internal build tooling
\cite{crowdstrike_rca}. \\[3pt]
Jul.~29 & Vendor connection metric reached approximately 99\% of the
comparison level at 17:00 PT, reported in the July 31, 16:38 UTC update
\cite{crowdstrike_recovery_hub}. \\[3pt]
Aug.~6 & RCA published, distinguishing implemented corrections
from fixes still scheduled for release \cite{crowdstrike_rca}. \\[3pt]
Aug.~9 & Scheduled general availability of sensor hotfixes for bounds
and count checks and provision of 21 inputs \cite{crowdstrike_rca}. \\[3pt]
Aug.~19 & Scheduled production release of additional Content Validator
checks \cite{crowdstrike_rca}. \\[3pt]
Sep.~24 & Written testimony reported new validation and runtime checks,
expanded tests, customer controls, and deployment rings as implemented,
without confirming the scheduled dates \cite{meyers_testimony}. \\
\bottomrule
\end{tabular}
\end{table*}

The RCA reports manual and automated Template Type testing, including
12 automated cases spanning supported Windows versions and a broad subset
of operational use cases. The content selected for those cases used a
regular-expression wildcard in field 21. Earlier production instances did
so as well \cite[pp.~2, 4]{crowdstrike_rca}. Consequently, successful
execution did not require a read of the missing input. The relevant
coverage gap was semantic---whether a field's value had to be accessed---not
simply insufficient operating-system diversity or load.

The July 19 update introduced a non-wildcard criterion for that field.
Windows hosts running sensor 7.11 or later that received the update during
the 04:09--05:27 UTC distribution window were exposed. At a subsequent
IPC notification, evaluation could reach the missing input and fault.
Linux and macOS hosts were outside the affected Channel File 291 path
\cite{crowdstrike_pir,crowdstrike_techdetails}. The distribution interval
does not establish that every affected host failed at the same time.

\subsection{From Input Mismatch to Kernel Fault}

The RCA's representative dump shows an array of 20 pointers to input
string structures. The requested zero-based index was 20
(\texttt{0x14}), one beyond the valid range. An unchecked array access
loaded a value from beyond that array; a later instruction dereferenced
the invalid pointer. The resulting
\texttt{PAGE\_FAULT\_IN\_NONPAGED\_AREA} bugcheck had code
\texttt{0x50} \cite[pp.~9--12]{crowdstrike_rca}.

This distinction matters for the corrective action. A null-pointer check
before the later dereference cannot establish that the earlier array
lookup was within bounds. A non-null value obtained from the wrong memory
location remains invalid. Microsoft's published disassembly shows such
a null check and the faulting read at \texttt{csagent+0xe14ed}, consistent
with the vendor's memory-safety explanation
\cite{microsoft_security_best_practices}. Neither that published analysis
nor this article establishes that all crashes had identical machine state.

Microsoft corroborates the read fault, not exploitability. The RCA
characterizes the defect as not involving an arbitrary memory write and
states that it was independently reviewed, but does not publish that
assessment \cite[p.~3]{crowdstrike_rca}. CISA's initial July 19 alert
relayed CrowdStrike's confirmation that the outage resulted from the
content update rather than malicious activity \cite{cisa_alert}.
The reliability failure is serious without requiring an unsupported
claim of exploitability.

\subsection{Distribution Reversion and Endpoint Recovery}

Reverting Channel File 291 stopped further distribution of the problematic
content. It did not automatically remove that content from every endpoint.
The PIR states that Channel Files were written to disk, and the remediation
hub notes that deprecated content could remain in the sensor directory
\cite{crowdstrike_pir,crowdstrike_recovery_hub}. A host that faulted on each
startup before fetching and applying replacement content could not recover
through the ordinary update loop alone.

The recovery options depended on how far a host could progress before
faulting. In its July 21, 09:32 UTC guidance, CrowdStrike recommended
rebooting on a wired network so that replacement content might arrive
before another crash \cite{crowdstrike_recovery_hub}. Meyers' written
testimony dates the introduction of automated remediation techniques to
July 22 \cite[p.~4]{meyers_testimony}. The hub's July 24, 03:35 UTC update
reports that the problematic file was placed on a cloud known-bad list
in US-1, US-2, and EU on July 23, without new sensor code or Channel Files;
Gov-1 and Gov-2 customers could request the classification. The vendor
reported possible automatic recovery of boot-looping hosts with strong
connectivity. The hub's remediation section separately directed hosts
that kept crashing and could not stay online to alternative repair
procedures \cite{crowdstrike_recovery_hub}.

Microsoft's recovery guidance offered Windows Preinstallation Environment
(WinPE) media and a Safe Mode route, with procedures for USB, network boot,
and virtual machines \cite{microsoft_recovery_tool}. These options show why
``manual recovery'' is too broad a description: some environments could
automate substantial parts of restoration, whereas others needed local
intervention. Likewise, it is incorrect to characterize recovery
environments as universally unable to use a network.

Disk encryption introduced another dependency. WinPE recovery could require
a BitLocker recovery key. The Safe Mode option could avoid key entry for
some Trusted Platform Module (TPM)-only configurations when a local
administrator account was available; TPM-plus-PIN configurations had
different requirements \cite{microsoft_recovery_tool}. The operational
issue was therefore access to the appropriate recovery credentials and
workflow, not an inherent incompatibility between encryption and repair.

\subsection{Observed Impact and Its Boundaries}

Microsoft's estimate of 8.5 million affected devices, fewer than 1\% of
all Windows machines, is the principal scale indicator
\cite{microsoft_weston_blog}. Its announcement does not describe an
estimation procedure. CISA repeated the estimate in its July 21,
09:45 EDT update, rather than providing independent confirmation
\cite{cisa_alert}. The figure should not be interpreted as a census derived
solely from Windows crash reports. Microsoft explicitly notes that the
reports it receives are a sampled subset of impacted devices, limited to
customers who share them
\cite{microsoft_security_best_practices}. The public record does not
provide a device-level distribution of crash counts, downtime, or repair
effort. Microsoft's published chart shows a July 19 peak and lower but
nonzero report counts through July 25; this separate temporal indicator
is not a recovery census \cite{microsoft_security_best_practices}.

CrowdStrike's July 31, 16:38 UTC measurement report states that Windows
sensor connections were approximately 99\% of the pre-incident comparison
level by July 29 at 17:00 PT, using a week-over-week comparison with
approximately 1\% normal variation \cite{crowdstrike_recovery_hub}.
The proxy compares sensor connections, not repaired hosts, repair effort,
or restored services. Meyers later characterized virtually all customer
systems as back up by July 29 \cite[p.~4]{meyers_testimony}. The published
measurement describes connections returning near the comparison level
about ten days after release reversion. It does not report individual-host
outage durations.

Operational effects continued after the 78-minute distribution window.
Delta reported approximately 7,000 flight cancellations over five days,
with approximately \$380 million in direct revenue impact, \$170 million
in non-fuel expense impact, and a \$50 million fuel-expense offset
\cite{delta_results}. These components describe Delta's reported financial
impact; they are neither global-loss estimates nor sensor restoration
times. Delta attributed the effects to the outage. In an
October 25 complaint, CrowdStrike disputed responsibility for the prolonged
disruption, alleging that Delta's response and IT infrastructure extended
its recovery \cite[pp.~1--3]{crowdstrike_delta_complaint}. The analysis
records the competing positions as of October 31, 2024, without
adjudicating them. Allocating recovery delay would require
organization-specific records separating endpoint repair from service
restoration.

CRS compiled reports of manual call handling in Phoenix and Portland after
computerized dispatch was affected, impaired or unavailable 911 calling
elsewhere, and jurisdictions reporting no 911 impact
\cite{congress_crs}. These reports establish exposure
of safety-relevant services; they do not quantify resulting injuries or
deaths. The FCA's qualitative assessment of UK regulated firms reported
varying operational effects and minimal consumer harm, based on engagement
with firms and incident notifications whose timing and completeness
varied \cite{fca_resilience}. This counterevidence limits claims of uniform
service disruption; it does not separate differences in endpoint exposure,
preparedness, or reporting.

CISA's initial July 19 alert reported observing threat actors using the
incident for phishing and other malicious activity; its July 20 and
July 24 updates relayed CrowdStrike reports of a malicious archive and
brand-impersonating domains \cite{cisa_alert}. Trusted recovery
communications were therefore part of the response, not merely
administrative details.

\begin{figure*}[t]
\centering
\includegraphics[width=\textwidth]{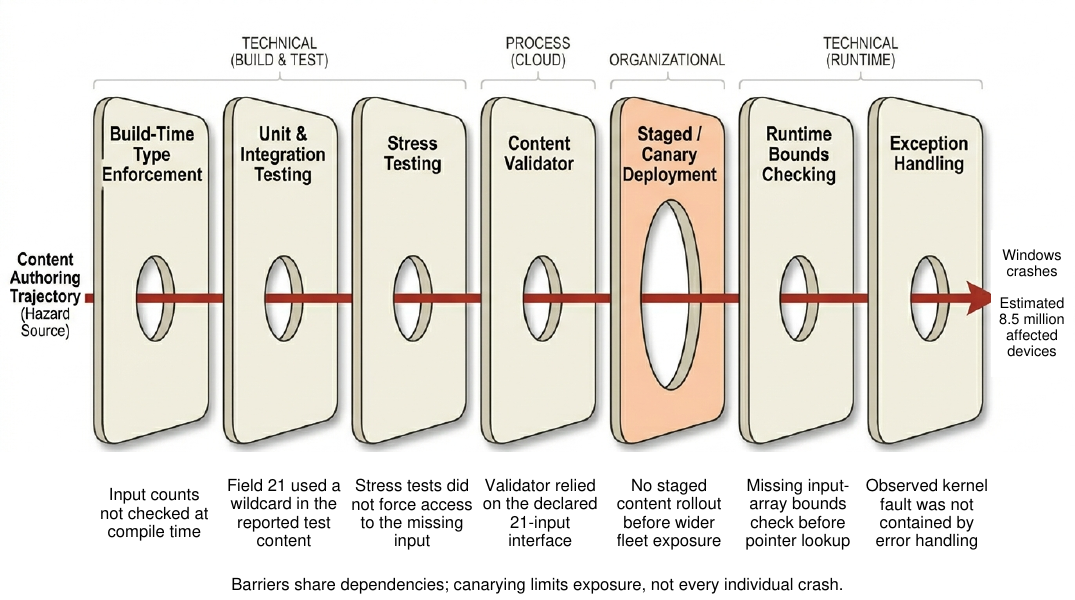}
\caption{Qualitative barrier view, informed by Reason's system approach
\cite{reason2000} and the incident reports
\cite{crowdstrike_rca,crowdstrike_pir}. Openings denote assurance gaps, not
measured probabilities. The barriers share dependencies; preventing a
fault, limiting fleet exposure, and restoring a host are different
outcomes. The device count is Microsoft's estimate
\cite{microsoft_weston_blog}. For the exception-handling barrier, the
effective boundary is before invalid access, not assumed recovery after
a kernel fault; see Section~\ref{sec:fault-containment}.}
\label{fig:swisscheese}
\end{figure*}

\begin{figure*}[t]
\centering
\includegraphics[width=\textwidth]{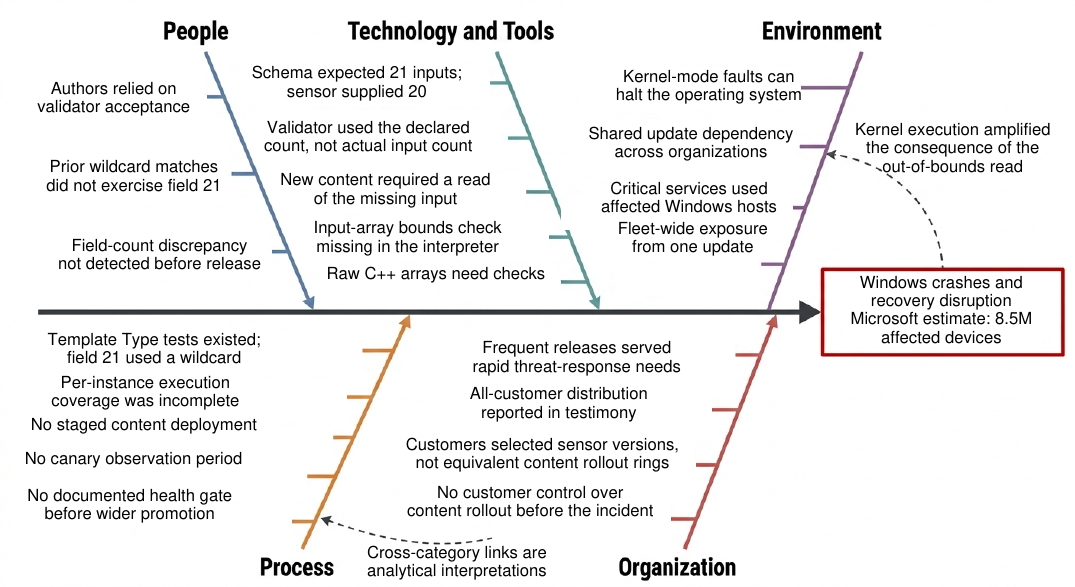}
\caption{Fishbone organization of reported technical and process findings
and their operational context
\cite{crowdstrike_rca,crowdstrike_pir,homeland_hearing,microsoft_security_best_practices,congress_crs,fca_resilience}.
The branches are categories, not quantitative rankings or independent
causes. Cross-category links express engineering interpretation; they do
not establish employee intent or a causal effect of market incentives.}
\label{fig:fishbone}
\end{figure*}

\section{Causal Analysis}
\label{sec:causal}

\subsection{Validation Shared an Unverified Assumption}

The content definition and runtime consumer disagreed about a basic
interface invariant: the number of available inputs. The Content Validator
evaluated an instance against the expected 21-field definition, while the
sensor supplied 20 inputs \cite{crowdstrike_rca}. The engineering
interpretation is that successful validation answered the wrong assurance
question: it established acceptability under the declared interface,
not compatibility with the deployed implementation.

This was not necessarily a second, unrelated corruption of the content.
The definition's 21-input expectation, which deployed sensor versions
7.11 and later did not meet, linked the authoring interface and validation
outcome. Repeating a check against that definition would not provide an
independent defense. The needed check was agreement with the consumer
that would execute the matching operation.

Binary certification did not remove this dependency. CrowdStrike reports
that Windows Hardware Quality Labs certification used the Channel Files
available when a sensor release was tested \cite[p.~8]{crowdstrike_rca}.
That procedure could not establish compatibility with every future
Template Instance. An unchanged, previously tested binary can follow a
different execution path when new data supplies different matching
criteria.

\subsection{Successful Tests Left the Relevant Path Unexercised}

The test record supports a narrower finding than ``configuration was not
tested.'' Template Type tests existed, covered supported Windows versions,
and exercised functional behavior. They failed to force evaluation of
the missing field. A wildcard that accepts any value can avoid the very
input access whose correctness the test is expected to establish
\cite[pp.~2--5]{crowdstrike_rca}.

Load testing and field-access testing answer different questions. A large
event volume can demonstrate throughput or stability along exercised paths
without demonstrating that every declared field can be read safely.
Likewise, several successful content releases provide weak evidence about
a field they never require the interpreter to inspect. The lesson is to
derive test obligations from the interface and matching semantics, not
from event volume or prior release success alone.

\subsection{Host-Level Fault Containment Was Insufficient}
\label{sec:fault-containment}

In the documented C++ input-array access, bounds safety depended on an
explicit check \cite[pp.~3, 7, 9--12]{crowdstrike_rca}. Its absence allowed
the interface defect to become an invalid memory access.
Kernel execution then made that access consequential for
host availability. The PIR describes exception handling intended to
contain problematic content, but the observed fault was not contained
\cite{crowdstrike_pir}. This does not show that every kernel exception is
unrecoverable; it shows that the protection in this path did not prevent
the bugcheck.

The preferred boundary is before invalid memory is accessed. A handler
cannot safely be assumed to recover from arbitrary corruption or an
unrecoverable kernel fault. Even with a bounds check, the application must
define what happens to the rejected detection rule, how the degraded
protection is reported, and whether a known-good configuration remains
available.

\subsection{Deployment Amplified Exposure; Recovery Had a Different Path}

In the committee's account, CrowdStrike's Adam Meyers stated that the
updates had been distributed to all customers in one session and that this
practice had since changed \cite{homeland_hearing}. Together with RCA
finding 6 \cite[p.~6]{crowdstrike_rca}, this documents the absence of an
exposure-limiting promotion stage for the incident update. It does not
establish that every device downloaded it simultaneously.

A staged rollout could have reduced exposure only if the initial cohort
exercised the failing path and an adverse signal stopped promotion.
Crashing hosts complicate both requirements. A release controller that
waits only for explicit sensor error reports can mistake missing reports
for success. Furthermore, stopping distribution cannot repair a host whose
ordinary update mechanism is no longer running. Prevention, containment,
and restoration are related, but none is a substitute for the others.

\subsection{Interpreting the Barrier and Fishbone Views}

Figure~\ref{fig:swisscheese} retains seven assurance opportunities spanning
build checks, testing, validation, deployment, and runtime handling.
The separation is useful for assigning verification work, but the
barriers are not independent: several relied on the same definition or
wildcard-bearing test content. Nor are all barriers necessary conditions
for an individual crash. A canary can constrain fleet exposure while
allowing a defect to crash a canary host. The diagram is therefore not a
fault tree and does not support multiplying barrier failure probabilities.

Figure~\ref{fig:fishbone} groups the same findings by engineering and
operational domain. Its organizational relationships are analytical
interpretations, not findings about budget decisions, competitive motives,
or individual knowledge. Frequent content delivery creates a need to manage
validation latency, but the public record does not establish that
commercial pressure caused a particular control to be omitted. Similarly,
shared exposure across organizations follows from a common update
dependency; a specific market-share estimate is unnecessary to establish
that mechanism.

\section{Engineering Controls and Their Limits}
\label{sec:controls}

Controls should be evaluated against the outcome they can change.
Interface checks prevent particular defects from reaching execution.
Runtime checks contain invalid inputs. Rollout gates limit initial
exposure. Recovery mechanisms restore service after those defenses fail.
Table~\ref{tab:controls} translates these distinctions into verification
criteria; the criteria are recommendations, not measurements performed
in this study.

\begin{table*}[t]
\caption{Recommended Controls, Verification Evidence, and Residual Risks}
\label{tab:controls}
\centering
\small
\begin{tabular}{@{}L{0.20\textwidth}L{0.44\textwidth}L{0.30\textwidth}@{}}
\toprule
Control and owner & Required verification evidence & Outcome and remaining limitation \\
\midrule
C1: Interface agreement.\newline Sensor and content teams.
& Gate each release against build-derived manifests for all supported
sensor versions; inject count, type, and optionality mismatches. Execute
Template Type/interpreter pairs when either changes, with focused
per-release tests.
& Rejects count, type, and optionality mismatches if supported-build
manifests are complete. Matching manifests do not prove that each input
is semantically correct. \\[3pt]
C2: Checked interpretation.\newline Sensor team.
& Test an index at the first invalid position, missing and mistyped inputs,
and invalid lengths. Reject the affected content before memory access;
verify host availability and an observable protection-status change.
& Contains invalid input. Error handling must not silently disable
protection or assume a kernel fault can be recovered after the fact. \\[3pt]
C3: Semantic execution tests.\newline Content assurance team.
& On Template Type or interpreter changes, force non-wildcard match and
non-match paths for every field, including update, restart, and rollback.
Execute every new instance; select its version/workload test matrix by
documented risk.
& Exposes untested access paths. Field coverage is necessary for this
mechanism but is not exhaustive behavioral coverage. \\[3pt]
C4: Health-gated promotion.\newline Release engineering.
& Inject a crashing update and a telemetry outage separately; promotion
must stop within a time and exposure budget justified against release
cadence and threat urgency. Attribute activation, heartbeat loss, and
reboots to the content generation under evaluation.
& Limits exposure if the cohort and signals are representative. A crash
can still affect the initial cohort; delayed faults may evade it. \\[3pt]
C5: Independent recovery.\newline Sensor, platform, and enterprise teams.
& Force a crash on every boot before connectivity; demonstrate an
authenticated local fallback decision before bad content is reactivated.
Also test out-of-band repair with encrypted disks, unavailable networks,
and inaccessible primary identity services.
& Reduces recovery dependence on failed hosts. Fallback protection,
tamper resistance, and recovery credentials remain constraints. \\[3pt]
C6: Service-level readiness.\newline Enterprise service owners.
& Restore an important service in an exercise that includes endpoint,
identity, management, and supplier dependencies; compare restoration time
with the service's agreed disruption tolerance.
& Addresses business-service recovery rather than device counts alone.
Redundant hosts may still share the same faulty update. \\
\bottomrule
\end{tabular}
\end{table*}

\subsection{Make the Consumer Contract Enforceable}

A shared, versioned interface should define field count, field types,
optional values, and compatibility rules. Generation or compile-time
assertions can prevent disagreement between the definition and input
provisioning code. The contract must also hold across supported sensor
versions: checking only the newest build is insufficient if older binaries
continue receiving new content.

A static compatibility gate should check every new content instance
against manifests generated from the input provisioning in every supported
sensor build, rather than from the template definition alone. These checks
cover count, type, and optionality without executing the full
version/workload matrix for every release. Dynamic field-access tests are
required whenever a Template Type or interpreter changes. Each content
release still needs execution testing, with supported versions and
workloads selected by documented risk. The wildcard-only field-21
restriction reported in the RCA illustrates the need to protect older
20-input sensors while corrected versions are deployed
\cite[p.~5]{crowdstrike_rca}.

The July 19 defect was a count mismatch, but a count check alone cannot
detect reordered fields, incorrect types, or an input with the wrong
meaning. Those cases require typed interfaces and tests of the actual
producer-consumer pairing. A schema compiler is a useful mechanism, not
proof of semantic correctness. This distinction is consistent with the
broader configuration-dependency problem described by Chen et al.\
\cite{chen2020_config_dependencies}.

Runtime checks remain necessary even when build-time agreement is enforced.
The interpreter should validate the requested index and the input
representation before access, reject incompatible content, and retain a
defined protection state. Rejection needs an explicit diagnostic and
operational status; silently treating invalid content as a successful
match or a healthy update would exchange a visible crash for an obscured
security failure.

\subsection{Test Field Access, Not Just Matching Outcomes}

For this defect class, a useful regression suite must force access to each
declared input. Non-wildcard match and non-match cases should be combined
with missing inputs, boundary indexes, empty values, and compatibility
checks against supported interpreter versions. Exercising realistic IPC
events matters because a content-only harness can supply the declared
21 inputs without reproducing the deployed sensor's 20-input interface.
The full field-access suite belongs to Template Type and interpreter
qualification; per-instance execution tests should target changed
criteria, boundary cases, and risk-selected version/workload combinations.
Static compatibility checks remain exhaustive across supported builds
for each release.

Field access coverage should be supplemented with update and recovery
sequences: loading new content during activity, restarting with persisted
content, reverting to a compatible baseline, and handling interrupted
updates. Fuzzing and fault injection can explore additional malformed
inputs and state transitions, but neither replaces a deterministic
regression for the documented 21-versus-20 failure.

\subsection{Gate Promotion on Observable Activation}

In the committee's account, Meyers described a cadence of 10--12 content
updates per day \cite{homeland_hearing}. Verification latency and deployment budgets must
therefore be designed against that cadence and the urgency of the threat
being addressed. Observation periods may overlap later releases; when
they do, health evidence must identify the content generation under test,
and promotion must wait for attributable evidence rather than aggregate
fleet stability.

Canary evaluation should be tied to content identity, sensor version,
activation time, and the workloads that exercise the changed rule.
Aggregate fleet health can conceal a failure in a small cohort. Cohorts
also need variation in Windows configuration, workload, restart behavior,
and recovery constraints, while avoiding an initial concentration in
irrecoverable or safety-critical systems.

A promotion interval must accommodate the time to activate the relevant
path, obtain trustworthy health evidence, and halt further distribution.
These quantities must be measured in the deployment environment; the
78-minute incident release-to-reversion interval is not a measurement of
canary detection latency. It includes an unspecified combination of
distribution, detection, investigation, and response.

The release controller should withhold promotion when required health
signals are missing, rather than interpreting silence as health.
Independent management or operating-system observations can help distinguish
sensor failure from an idle workload, although shared network or identity
failures can affect those channels too. Acknowledged content delivery is
not evidence that the matching path ran successfully. Google's canary
guidance similarly makes evaluation and integration with release decisions
essential parts of the mechanism \cite{google_canary}.

Customer controls should permit staged content activation with explicit
limits, exception handling, and visibility into the resulting protection
state. Delaying detection content can increase exposure to active threats.
The operating policy must therefore specify who can approve an urgent
override, what evidence is required, and how the exception is recorded;
neither unrestricted deferral nor unconditional global deployment resolves
the tradeoff. At the customer level, the FCA advised regulated firms to
consider phasing releases across user groups, and observed that firms
with mapped third- and nth-party relationships understood their exposure
quickly \cite{fca_resilience}.

\subsection{Design Recovery Outside the Failed Update Path}

A local, authenticated last-known-good content generation could provide a
recovery option when a new generation crashes on every boot before the
sensor gains connectivity.
This is a proposed design direction, not a commitment attributed to the
PIR. Its implementation would require durable content identity, atomic
activation, a boot-health decision independent of the failing interpreter,
and protection against an attacker deliberately inducing fallback.
The decision must precede reactivation of the suspect content, rather than
depend on a download completing before the next fault. CrowdStrike's
July 23 known-bad classification is a vendor-reported precedent for remote
suppression without new sensor code, but its connectivity requirement
does not establish an equivalent local recovery mechanism
\cite[Jul.~24 update]{crowdstrike_recovery_hub}.

Fallback is not automatically ``fail-safe.'' Retaining older content or
quarantining a rule preserves some availability at the cost of potentially
reduced protection. The permitted fallback duration, administrator
notification, reactivation conditions, and response to persistent failure
must follow the host's security and service requirements. A general-purpose
reboot counter cannot by itself attribute a crash to a particular content
generation.

Enterprises also need tested out-of-band procedures: accessible recovery
keys, appropriate local credentials, trusted boot media, and recovery
infrastructure that does not share all dependencies with the impaired
fleet. Microsoft's documented WinPE, Safe Mode, network-boot, and
virtual-machine options illustrate why a single recovery procedure cannot
be assumed to fit every deployment \cite{microsoft_recovery_tool}.
The vendor hub also documents bootable recovery images, repair of detached
virtual-machine disks, and rollback to snapshots from before July 19,
04:09 UTC, and links Intel vPro and Absolute guidance
\cite{crowdstrike_recovery_hub}. These are environment-specific options,
not evidence that every deployment had an independent management path.

\subsection{Reduce Privilege Without Assuming a Cost-Free Migration}

Moving suitable content interpretation into an isolated user-mode component
could reduce the host-level consequence of an interpreter failure. A
memory-safe implementation could remove some unsafe access paths.
Neither change is sufficient by itself: process failure, unchecked foreign
interfaces, resource exhaustion, or incorrect matching logic can still
degrade protection.

The feasibility of partitioning depends on callback timing, inline
enforcement, tamper resistance, data transfer, and restart behavior.
Microsoft's discussion of security-tool integration describes both the
reasons for kernel participation and opportunities to reduce it
\cite{microsoft_security_best_practices}. The public evidence does not
establish that a complete, behaviorally equivalent user-mode replacement
was available before the incident. A defensible migration would identify
the minimum privileged component and measure both security effectiveness
and availability.

\subsection{Distinguish Reported Remediation From Recommendations}

The August 6 RCA reports several distinct corrective actions. A
compile-time input-count check entered internal build tooling on July 27;
the vendor also reported checking all other Template Types and platforms
without finding another count discrepancy. On July 25, the interpreter
received both bounds checks and a check that supplied and expected input
counts matched. The sensor code was changed to supply the intended 21
inputs. Backports of these sensor corrections to versions 7.11 and later
were scheduled for general availability by August 9
\cite[pp.~3--4]{crowdstrike_rca}.

The report also describes completed Channel 291 fuzz testing, expanded
non-wildcard field tests, and testing of every new Template Instance.
Restricting field 21 to wildcard criteria protected unpatched 20-input
sensors: it was a compatibility guard of the kind required by C1, not
a declaration that 21 inputs were intrinsically incorrect. Additional
validator checks were scheduled for August 19. Deployment rings,
customer selection of where and when content would be delivered, and two
third-party reviews were reported as introduced or started
\cite[pp.~4--6]{crowdstrike_rca}.

Meyers' September 24 written testimony reports new validation and runtime
checks, expanded field tests, customer controls, and deployment rings as
implemented, and the third-party reviewers as engaged
\cite[pp.~4--5]{meyers_testimony}. It gives no completion dates, review
results, or fleet-wide effectiveness measurements, leaving the August
deadlines and individual endpoint coverage unverified. Controls C1--C4
overlap these vendor-reported changes; the local recovery design and
service-level exercises are broader recommendations requiring their own
implementation and evaluation.

\section{Operational Responsibility and Limits}
\label{sec:discussion}

\subsection{Allocate Responsibility by Control}

CrowdStrike controlled the defective interface, content validation, sensor
error handling, and distribution policy. It therefore held the principal
opportunities to prevent the defect and limit initial exposure. Microsoft
controlled the operating-system interfaces and platform recovery
capabilities, while enterprises controlled deployment architecture,
credential availability, and business-continuity preparation. These are
engineering responsibilities, not findings of legal liability.

Customer selection of sensor versions did not provide equivalent control
over Rapid Response Content. It would therefore be misleading to assign
customers responsibility for staging a content update through a mechanism
they did not have. Conversely, a vendor's primary responsibility for the
trigger does not eliminate the value of customer-side resilience.
The FCA observed that service mapping supported restoration priorities;
it separately reported benefits from rehearsed scenarios and faster
stakeholder communication with tested strategies \cite{fca_resilience}.
These observations support preparedness measures,
but do not quantify their causal effect on recovery time.

Redundancy also needs to be assessed against the failure source. Two hosts
in different facilities can share one content pipeline, sensor version,
identity service, or recovery-key repository. They are not independent
with respect to those dependencies. Diversification can reduce common
exposure, but additional products and operating environments also increase
integration and support costs. The relevant objective is demonstrable
service continuity, not diversity for its own sake.

\subsection{Use Standards as Guidance, Not Retrospective Certification}

The NIST Secure Software Development Framework, version 1.1, provides
relevant practices for security requirements and design (PW.1), code
review or analysis (PW.7), executable testing (PW.8), and analysis of root
causes to prevent recurrence (RV.3) \cite{nist_ssdf}. Applied here, they
support explicit interface requirements, checked interpretation, test
obligations, and a feedback loop from incidents to engineering controls.
The framework does not prescribe a particular canary size or establish
that a rollback process meets a service's availability requirements.

IEC 61508 addresses functional safety of electrical, electronic, and
programmable electronic safety-related systems \cite{iec61508}.
Its hazard-oriented perspective is useful when assessing how a security
component's loss of availability could affect a safety-related service.
However, this article does not classify Falcon as an IEC 61508 safety
system, assign a safety integrity level, or assess compliance. Such claims
would require a defined application safety function and substantially
more evidence than an outage reconstruction supplies.

Professional accountability does not require speculation about individual
intent. The public sources do not establish what particular employees knew
before release, and absence of such evidence neither proves misconduct
nor establishes that every decision met professional obligations. The ACM
Code of Ethics emphasizes avoiding harm, evaluating system risks, and
taking special care of infrastructure on which society depends
\cite{acm_code_ethics}. Those obligations support scrutiny of the assurance
process and its consequences without inventing a narrative of deliberate
negligence.

\subsection{Limits of the Counterfactuals}

A correct input-count check would reject this specific mismatch, and a
correctly implemented bounds check would prevent this specific
out-of-bounds access. These are mechanism-level conclusions, conditional
on checking the actual interface and defining a safe rejection path.
They are not evidence that all possible content defects would be prevented.

Claims about canary effectiveness are less determinate. A limited cohort
could expose the defect before wider promotion, but only if it activated
the relevant path, produced usable evidence, and caused distribution to
stop in time. The available sources do not disclose the necessary
activation distributions, telemetry delays, cohort composition, or
the number of endpoints simultaneously receiving or activating a given
content generation. The reported release cadence does not supply those
quantities. A numerical estimate of avoided crashes, an optimal ring
percentage, or a guaranteed detection interval would therefore be
unsupported.

Recovery improvements have similar limits. Reliable access to repair media
and credentials can remove specific obstacles, but restoration of an
endpoint does not ensure immediate restoration of a dependent business
service. Neither the Microsoft device estimate nor Delta's service impact
provides the paired host-and-service measurements needed to quantify that
relationship. Delta's causal attribution is also contested; its aggregate
financial figures cannot allocate delay among endpoint repair and
organization-specific restoration dependencies.

Finally, the technical reconstruction relies heavily on the vendor for
implementation and test-process details. Microsoft's analysis corroborates
the fault, not all internal explanations. The diagrams simplify a
proprietary architecture, and no experiments, interviews, or source-code
inspection were performed for this article. Generalization is therefore
limited to the identified mechanisms and assurance questions, not a
measured failure rate for EDR products as a class.

Later platform commitments are outside the evidence window. Microsoft's
November 19, 2024 Windows Resiliency Initiative announcements included
recovery for nonbooting PCs, gradual security-product updates, and
user-mode integration \cite{microsoft_resiliency}. Their delivery and
effectiveness are not assessed here.

\section{Conclusion}

Channel File 291 exposed a mismatch between a declared detection interface
and the inputs its runtime consumer actually received. Wildcard-bearing
tests and earlier content did not require the missing input to be read;
a later criterion did. Missing bounds validation turned that discrepancy
into a kernel fault. Unstaged distribution widened exposure, while
persistent content and impaired hosts complicated restoration. Some
hosts could recover through network-assisted remediation; those that
could not sustain the necessary connectivity required another path.

The engineering lesson is not that rapid security updates should stop,
or that labeling content as data makes it harmless. Assurance must follow
the behavior an update can activate. For privileged consumers, this
requires enforceable compatibility contracts, tests that exercise the
declared interface, explicit handling of rejected content, health-gated
deployment, and a recovery path that remains usable when the normal update
path fails.

These controls address different outcomes and share some dependencies.
Their effectiveness should be demonstrated separately: prevention of the
documented input fault, containment of a defective release, and restoration
of the affected service. Treating those as distinct engineering obligations
is more useful than attributing the outage to one bad file or assuming
that distribution reversion completed recovery.

\IEEEtriggeratref{10}
\bibliographystyle{IEEEtran}
\bibliography{refs}

\begin{thebibliography}{10}
\providecommand{\url}[1]{#1}
\csname url@samestyle\endcsname
\providecommand{\newblock}{\relax}
\providecommand{\bibinfo}[2]{#2}
\providecommand{\BIBentrySTDinterwordspacing}{\spaceskip=0pt\relax}
\providecommand{\BIBentryALTinterwordstretchfactor}{4}
\providecommand{\BIBentryALTinterwordspacing}{\spaceskip=\fontdimen2\font plus
\BIBentryALTinterwordstretchfactor\fontdimen3\font minus
  \fontdimen4\font\relax}
\providecommand{\BIBforeignlanguage}[2]{{%
\expandafter\ifx\csname l@#1\endcsname\relax
\typeout{** WARNING: IEEEtran.bst: No hyphenation pattern has been}%
\typeout{** loaded for the language `#1'. Using the pattern for}%
\typeout{** the default language instead.}%
\else
\language=\csname l@#1\endcsname
\fi
#2}}
\providecommand{\BIBdecl}{\relax}
\BIBdecl

\bibitem{crowdstrike_rca}
\BIBentryALTinterwordspacing
{CrowdStrike}, ``External technical root cause analysis---{Channel File 291},''
  CrowdStrike, Tech. Rep., Aug.~6, 2024, {Accessed}: Oct. 22, 2026. [Online].
  Available:
  \url{https://www.crowdstrike.com/wp-content/uploads/2024/08/Channel-File-291-Incident-Root-Cause-Analysis-08.06.2024.pdf}
\BIBentrySTDinterwordspacing

\bibitem{microsoft_weston_blog}
\BIBentryALTinterwordspacing
D.~Weston. (2024, Jul.) Helping our customers through the {CrowdStrike} outage.
  Microsoft. Accessed: Oct. 22, 2026. [Online]. Available:
  \url{https://blogs.microsoft.com/blog/2024/07/20/helping-our-customers-through-the-crowdstrike-outage/}
\BIBentrySTDinterwordspacing

\bibitem{crowdstrike_pir}
\BIBentryALTinterwordspacing
{CrowdStrike}. (2024, Jul.) Falcon content update preliminary post incident
  report. Updated Jul. 25, 2024. Accessed: Oct. 22, 2026. [Online]. Available:
  \url{https://www.crowdstrike.com/en-us/blog/falcon-content-update-preliminary-post-incident-report/}
\BIBentrySTDinterwordspacing

\bibitem{crowdstrike_techdetails}
\BIBentryALTinterwordspacing
------. (2024, Jul.) Technical details: {Falcon} update for {Windows} hosts.
  Accessed: Oct. 22, 2026. [Online]. Available:
  \url{https://www.crowdstrike.com/en-us/blog/falcon-update-for-windows-hosts-technical-details/}
\BIBentrySTDinterwordspacing

\bibitem{microsoft_security_best_practices}
\BIBentryALTinterwordspacing
D.~Weston. (2024, Jul.) {Windows} security best practices for integrating and
  managing security tools. Microsoft. Accessed: Oct. 22, 2026. [Online].
  Available:
  \url{https://www.microsoft.com/en-us/security/blog/2024/07/27/windows-security-best-practices-for-integrating-and-managing-security-tools/}
\BIBentrySTDinterwordspacing

\bibitem{microsoft_recovery_tool}
\BIBentryALTinterwordspacing
{Microsoft}. (2024, Jul.) New recovery tool to help with {CrowdStrike} issue
  impacting {Windows} endpoints. Updated Jul. 31, 2024. Accessed: Oct. 22,
  2026. [Online]. Available:
  \url{https://techcommunity.microsoft.com/blog/intunecustomersuccess/new-recovery-tool-to-help-with-crowdstrike-issue-impacting-windows-endpoints/4196959}
\BIBentrySTDinterwordspacing

\bibitem{crowdstrike_recovery_hub}
\BIBentryALTinterwordspacing
{CrowdStrike}. (2024) Falcon content update remediation and guidance hub.
  Statement-level update dates identified in text. Accessed: Oct. 22, 2026.
  [Online]. Available:
  \url{https://www.crowdstrike.com/falcon-content-update-remediation-and-guidance-hub/}
\BIBentrySTDinterwordspacing

\bibitem{homeland_hearing}
\BIBentryALTinterwordspacing
{U.S. House Committee on Homeland Security}. (2024, Sep.) Committee examines
  {CrowdStrike} processes in first congressional hearing on the disastrous
  {July} global {IT} outage. Account of Sep. 24, 2024 testimony. Accessed: Oct.
  22, 2026. [Online]. Available:
  \url{https://homeland.house.gov/2024/09/26/icymi-committee-examines-crowdstrike-processes-in-first-congressional-hearing-on-the-disastrous-july-global-it-outage/}
\BIBentrySTDinterwordspacing

\bibitem{meyers_testimony}
\BIBentryALTinterwordspacing
A.~Meyers, ``Testimony of {Adam Meyers},'' written statement of the {Senior
  Vice President, Counter Adversary Operations, CrowdStrike Holdings, Inc.},
  hearing of the {Subcommittee on Cybersecurity and Infrastructure Protection},
  {Committee on Homeland Security}, {U.S. House of Representatives},
  Washington, DC, USA, Sep.~24, 2024, {Accessed}: Oct. 22, 2026. [Online].
  Available:
  \url{https://homeland.house.gov/wp-content/uploads/2024/09/2024-09-24-HRG-CIP-Testimony-Meyers.pdf}
\BIBentrySTDinterwordspacing

\bibitem{congress_crs}
\BIBentryALTinterwordspacing
C.~L. Pechtol and J.~C. Gallagher, ``{IT} outage from {CrowdStrike's} update:
  Impacts to certain public safety systems and considerations for {Congress},''
  Congressional Research Service, {CRS In Focus} IF12717, Jul.~26, 2024,
  {Accessed}: Oct. 22, 2026. [Online]. Available:
  \url{https://www.congress.gov/crs_external_products/IF/PDF/IF12717/IF12717.1.pdf}
\BIBentrySTDinterwordspacing

\bibitem{delta_results}
\BIBentryALTinterwordspacing
{Delta Air Lines}. (2024, Oct.) {Delta Air Lines} announces {September} quarter
  2024 financial results. Accessed: Oct. 22, 2026. [Online]. Available:
  \url{https://ir.delta.com/news/news-details/2024/Delta-Air-Lines-Announces-September-Quarter-2024-Financial-Results/default.aspx}
\BIBentrySTDinterwordspacing

\bibitem{fca_resilience}
\BIBentryALTinterwordspacing
{Financial Conduct Authority}. (2024, Oct.) {CrowdStrike} outage: Lessons for
  operational resilience. Accessed: Oct. 22, 2026. [Online]. Available:
  \url{https://www.fca.org.uk/firms/operational-resilience/crowdstrike-outage-lessons-operational-resilience}
\BIBentrySTDinterwordspacing

\bibitem{crowdstrike_delta_complaint}
\BIBentryALTinterwordspacing
{CrowdStrike, Inc.}, ``Complaint, {CrowdStrike, Inc. v. Delta Air Lines,
  Inc.}'' U.S. District Court, Northern District of Georgia, No.
  1:24-cv-04904-TWT, Doc. 1, Oct.~25, 2024, {Accessed}: Oct. 22, 2026.
  [Online]. Available:
  \url{https://storage.courtlistener.com/recap/gov.uscourts.gand.335602/gov.uscourts.gand.335602.1.0.pdf}
\BIBentrySTDinterwordspacing

\bibitem{cisa_alert}
\BIBentryALTinterwordspacing
{Cybersecurity and Infrastructure Security Agency}. (2024, Jul.) Widespread
  {IT} outage due to {CrowdStrike} update. Updated Aug. 6, 2024;
  statement-level dates identified in text. Accessed: Oct. 22, 2026. [Online].
  Available:
  \url{https://www.cisa.gov/news-events/alerts/2024/07/19/widespread-it-outage-due-crowdstrike-update}
\BIBentrySTDinterwordspacing

\bibitem{chen2020_config_dependencies}
Q.~Chen, T.~Wang, O.~Legunsen, S.~Li, and T.~Xu, ``Understanding and
  discovering software configuration dependencies in cloud and datacenter
  systems,'' in \emph{Proc. 28th ACM Joint European Software Engineering Conf.
  and Symp. Foundations of Software Engineering (ESEC/FSE)}, Nov. 2020, pp.
  362--374, doi:
  \href{https://doi.org/10.1145/3368089.3409727}{10.1145/3368089.3409727}.

\bibitem{google_canary}
\BIBentryALTinterwordspacing
A.~Warner and {\v{S}}.~Davidovi{\v{c}}, ``Canarying releases,'' in \emph{The
  Site Reliability Workbook: Practical Ways to Implement {SRE}}, B.~Beyer,
  N.~R. Murphy, D.~K. Rensin, K.~Kawahara, and S.~Thorne, Eds.\hskip 1em plus
  0.5em minus 0.4em\relax O'Reilly Media, 2018, ch.~16, with A. Hidalgo, B.
  Beyer, K. Smith, and M. Duftler. Accessed: Oct. 22, 2026. [Online].
  Available: \url{https://sre.google/workbook/canarying-releases/}
\BIBentrySTDinterwordspacing

\bibitem{reason2000}
J.~Reason, ``Human error: Models and management,'' \emph{BMJ}, vol. 320, no.
  7237, pp. 768--770, Mar. 2000, doi:
  \href{https://doi.org/10.1136/bmj.320.7237.768}{10.1136/bmj.320.7237.768}.

\bibitem{nist_ssdf}
M.~Souppaya, K.~Scarfone, and D.~Dodson, ``Secure software development
  framework ({SSDF}) version 1.1: Recommendations for mitigating the risk of
  software vulnerabilities,'' National Institute of Standards and Technology,
  Tech. Rep. NIST SP 800-218, Feb. 2022, doi:
  \href{https://doi.org/10.6028/NIST.SP.800-218}{10.6028/NIST.SP.800-218}.

\bibitem{iec61508}
\emph{Functional Safety of Electrical/Electronic/Programmable Electronic
  Safety-Related Systems---Part 1: General Requirements}, {IEC 61508-1:2010},
  Ed.~2.0, International Electrotechnical Commission, Geneva, Switzerland,
  2010.

\bibitem{acm_code_ethics}
\BIBentryALTinterwordspacing
{Association for Computing Machinery}. (2018, Jul.) {ACM} code of ethics and
  professional conduct. Accessed: Oct. 22, 2026. [Online]. Available:
  \url{https://www.acm.org/code-of-ethics}
\BIBentrySTDinterwordspacing

\bibitem{microsoft_resiliency}
\BIBentryALTinterwordspacing
D.~Weston. (2024, Nov.) {Windows} security and resiliency: Protecting your
  business. Microsoft. Cited only to delimit follow-up work outside the
  evidence window. Accessed: Oct. 22, 2026. [Online]. Available:
  \url{https://blogs.windows.com/windowsexperience/2024/11/19/windows-security-and-resiliency-protecting-your-business/}
\BIBentrySTDinterwordspacing

\end{thebibliography}

\end{document}